\pdfoutput=1
\documentclass[12pt]{article}
\usepackage{amsmath,amssymb}
\usepackage{graphicx}
\usepackage{hyperref}
\numberwithin{equation}{section}
\def\be{\begin{equation}}
\def\ee{\end{equation}}

\def\bea{\begin{eqnarray}}
\def\eea{\end{eqnarray}}

\title{Extended Local Infrared Cutoff for the Holographic Density}
\author{L.N. Granda\thanks{luis.granda@correounivalle.edu.co}\\{\small\it Departamento de Fisica, Universidad del Valle, A.A. 25360 Cali, Colombia}}

\date{}
\begin{document}
\maketitle

\begin{abstract}
\noindent We propose an extended local infrared cutoff for the holographic density that leads to linear cosmological equations and gives exact analytical solutions. The model allows to remove the residual matter and radiation-like terms of the local construction based on $H$ and $\dot{H}$, resulting in a clear separation between the standard cosmological fluids and the holographic sector. The homogeneous solution generates
characteristic modes that give rise to different cosmological behaviors. The extension includes derivatives up to third order, where two of the characteristic modes can modify the pre-recombination expansion and reduce the sound horizon while remaining subdominant at BBN. A slightly phantom mode provides an additional late-time geometrical freedom that can further increase the CMB-compatible value of $H_0$. Optimizations using BAO constraints yield values of $H_0$ in the approximate range $70.5-72.4\ {\rm km\;s^{-1}\;Mpc^{-1}}$. 
\end{abstract}

\section{Introduction}
                   
\noindent The discovery of the accelerated expansion of the Universe from Type Ia supernovae  \cite{Riess1998, Perlmutter1999}, supported by measurements of the cosmic microwave background (CMB), baryon acoustic oscillations (BAO) and large-scale structure \cite{Planck2020}-\cite{eBOSS2021}, established the $\Lambda$CDM model as the standard description of the cosmological background \cite{Weinberg1989}-\cite{Frieman2008}. These observations also suggests that the universe is spatially flat, and consists of about $\sim 25-27\%$ of dark matter, $\sim 5\%$ of baryonic matter and $\sim 68-70\%$ of homogeneously distributed dark energy. Despite the abundance of the dark energy component its nature remains unknown and it is one of the fundamental problems of modern science. Although the $\Lambda$CDM model fits observational data very well, it suffers from the well-known fine-tuning problem, as well as the coincidence problem, which has motivated the study of alternative dynamical models to explain the vast amount of observational data from both the early and late universe.
Added to this is the current problem of the $H_0$ tension, which has reignited the search for dark energy models that attempt to explain this discrepancy, potentially pointing toward the existence of new physics beyond the Standard Model.\\
Among the dynamic models proposed to explain dark energy, holographic models are particularly interesting because they incorporate certain concepts from the quantum theory and gravity in what is known as the holographic principle. In  \cite{Bekenstein1973} it was established 
that gravitational entropy scales with the horizon area, motivating the holographic principle developed by 't Hooft and Susskind  \cite{Hooft1993, Susskind1995}. In \cite{Cohen1999} it was suggested that an effective quantum field theory in a region of characteristic size \(L\) must satisfy an ultraviolet-infrared relation preventing black hole formation. Thus, if $L$ is the size of the region and \(\rho\) its density, then \(\rho\le M_p^2L^{-2}\). 
Saturation of this bound leads to a holographic energy density of the form
$$
\rho_H=3c^2M_P^2L^{-2},
$$
so that the cosmological dynamics is essentially determined by the choice of the infrared (IR) cutoff $L$.\\
A natural choice for the infrared cutoff would be the Hubble radius, but it does not produce the effect of accelerated expansion in the standard non-interacting scenario \cite{Hsu2004}. The particle horizon and future event horizon were introduced in Ref. \cite{Li2004}. While the particle horizon leads to an equation of state (EOS) \(w>-1/3\) and cannot produce accelerated expansion, the future event horizon overcomes this difficulty and has generated a broad class of holographic dark-energy models \cite{Huang2005}-\cite{Li2011}. However, its dependence on the future expansion history motivates the search for local alternatives. The Ricci curvature provides one such possibility \cite{Gao2009}, while a local cutoff constructed from linear combination of \(H^2\) and \(\dot H\) was introduced in Refs. \cite{Granda2008,Granda2009}. This model depends exclusively on quantities evaluated at a given cosmic time, avoiding non-locality, and describing late-time accelerated expansion. Its cosmological properties and different extensions have been investigated in Refs.  \cite{Xu2009}-\cite{Malekjani2013}.\\
\noindent The holographic framework has also been generalized through modified entropy relations and alternative gravitational proposals  \cite{Nojiri2011}-\cite{Saridakis2020}. 
Generalized entropies have motivated new holographic dark-energy scenarios  \cite{Nojiri2017}-\cite{Barrow2020}, while the works of Nojiri, Odintsov and collaborators have established broader connections between holographic descriptions, generalized entropy, modified gravity and dark-energy cosmology  \cite{Nojiri2011}-\cite{Nojiri2017}, \cite{Nojiri2017a, Nojiri2021}. These developments suggest that the infrared cutoff may encode a richer local structure associated with the cosmological dynamics.\\
In the local infrared cutoff \cite{Granda2008}, the holographic sector of the solution to the Friedmann equation contains terms proportional to \(a^{-3}\) and \(a^{-4}\) that scale exactly like matter and radiation. Although these terms can be absorbed into effective density parameters, the holographic sector is not completely separated from the physical cosmological fluids. This residual presence of matter and radiation terms in the holographic density suggests that this cutoff is incomplete and can
be extended to a more general construction. In this paper, we propose such a construction, while attempting to preserve the locality and analytical simplicity of the model \cite{Granda2008}.\\
The new terms containing up to third-order derivatives are organized in such a way that the resulting Friedmann equation retains the structure of a linear differential equation with constant coefficients. This leads to a local infrared cutoff capable of describing a broader class of cosmological evolutions while preserving the original holographic motivation. Higher derivatives of the Hubble parameter arise in modified-gravity and higher-order cosmological theories \cite{Felice2010}-\cite{Nojiri2007}, and related extensions have also been considered in holographic cosmology \cite{Gong2010}.\\
\noindent In the present work we use the additional freedom arising in the extended cutoff, to constraint the parameters of the model so that the coefficients of the matter and radiation terms in the Friedmann solution remain equal to their standard physical values. Therefore, the residual holographic contributions that scale as \(a^{-3}\) and \(a^{-4}\) are eliminated, leaving a clear separation between matter, radiation and holographic sector. The holographic sector becomes entirely described by the characteristic modes of the third-order homogeneous equation and leads to a richer class of dynamical dark-energy solutions while preserving the standard matter and radiation-dominated epochs.\\
\noindent On the other hand, the mentioned $H_0$ tension, which is the persistent discrepancy between early and late time determinations of the Hubble constant, has motivated extensive investigations \cite{Riess2022}-\cite{Valentino2021}, particularly of scenarios in which the pre-recombination expansion rate is modified and the sound horizon is reduced \cite{Valentino2021}-\cite{Sakstein2020}. The higher-order holographic modes generated by the present construction provide a natural realization of this possibility without the introduction of early dark energy terms, since some of them scale between matter and radiation and may therefore enhance the expansion rate around recombination and matter-radiation equality while becoming progressively less important toward BBN. \\
For our specific extension that contains derivatives up to third order, we derive the corresponding cosmological equation, obtain its exact solution and classify the characteristic modes. We then investigate whether some of these modes can reduce the sound horizon and contribute to alleviating the  \(H_0\) tension. We also consider a slightly phantom characteristic mode and show that it provides an additional late-time geometrical freedom, although the resulting amplification of  \(H_0\) is significantly restricted once BAO distances are taken into account. The present model is analyzed at a background level. A complete study requires cosmological perturbations and a joint analysis of CMB, BAO, supernova and large-scale-structure observations and is beyond the scope of the present study.\\
\noindent The paper is organized as follows. In Section 2 we introduce the extended local infrared cutoff, derive the cosmological field equation and obtain the conditions for exact matter-radiation separation. We also present the analytical solution and classify the relevant characteristic modes. In Section 3 we study the early-time modes and their effect on the sound horizon. In Section 4 we analyze the additional geometrical amplification produced by a slightly phantom mode and its compatibility with representative BAO constraints. Our conclusions are presented in Section 5.

\section{Extended Local Infrared Cutoff}
The holographic principle applied to cosmology leads to the proposal of an holographic energy density proportional 
to the inverse square of an infrared cutoff. In \cite{Granda2008} the infrared cutoff was constructed from local geometric quantities leading to the holographic density
\be\label{eq2.1}
\rho=3M_p^2\left(\alpha H^2+\beta\dot{H}\right)
\ee
where \(M_p=(8\pi G)^{-1/2}\) is the reduced Planck mass.\\
The solution of the Friedmann equation with matter and radiation terms is
\be\label{eq2.1a}
H^2=H_0^2\left[A_m\Omega_{m0}e^{-3x}+A_r\Omega_{r0}e^{-4x}+\Omega_{\Lambda}e^{-2(\alpha-1)x/\beta}\right]
\ee
$$\Omega_{m0}=\frac{\rho_{m0}}{3M_p^2 H_0^2},\;\;\; \Omega_{r0}=\frac{\rho_{r0}}{3M_p^2 H_0^2},$$ are the matter and radiation density parameters, and  
\be\label{eq2.1b}
A_m=\frac{2}{3\beta-2\alpha+2},\;\; A_r=\frac{1}{2\beta-\alpha+1}
\ee
The coefficients \(A_m\) and \(A_1\) are different form unity cannot both be set equal to $1$ simultaneously. Therefore, the holographic component is not completely decoupled from the standard cosmological fluids and the matter and radiation sectors entering the expansion history are effectively mixed with the holographic contribution.\\
To avoid this problem we propose a more general local infrared cutoff that leads to the following holographic density
\be\label{eq2.2b}
\rho_{\cal{H}}=3M_p^2\left[\alpha H^2+\beta \dot{H}+\gamma\frac{\ddot{H}}{H}+\eta\frac{1}{H}\frac{d}{dt}\left(\frac{\ddot{H}}{H}\right)\right],
\ee
where $\cal{H}$ stands for "holographic". Working in the \(x=\ln a\) variable and defining \(H^2(x)=E(x)\) the derivative terms in (\ref{eq2.2b}) become
$$\dot{H}=\frac{dH}{dt}=\frac{1}{2}\frac{dE}{dx},$$
$$\frac{\ddot{H}}{H}=\frac{1}{H}\frac{d}{dt}\left(\frac{dH}{dt}\right)=\frac{1}{2}\frac{d^2E}{dx^2},$$
$$\frac{1}{H}\frac{d}{dt}\left(\frac{\ddot{H}}{H}\right)=\frac{1}{H}\frac{d}{dt}\left[\frac{1}{H}\frac{d}{dt}\left(\frac{dH}{dt}\right)\right]=\frac{1}{2}\frac{d^3E}{dx^3}.$$
Then, the Friedmann equation becomes a linear differential equation 
\be\label{eq2.3}
E(x)=H_0^2\left(\Omega_{m0}e^{-3x}+\Omega_{r0}e^{-4x}\right)+\alpha E(x)+\frac{1}{2}\beta\frac{dE}{dx}+\frac{1}{2}\gamma\frac{d^2E}{dx^2}+\eta\frac{1}{2}\frac{d^3E}{dx^3}.
\ee
Thus, the cosmological evolution is governed by a linear third-order differential equation with constants coefficients. The linearity allows the complete analytical solution of the cosmological background while maintaining a transparent interpretation of the homogeneous and particular solutions. The late-time evolution is controlled by the characteristic exponents which are determined by a cubic polynomial. The term with second derivative has been considered in \cite{Chen2009} for the study of late time cosmology. The discussion of this case, corresponding to $\eta=0$, will be given bellow.\\
\\
Note that mathematically we can generalize to an $n$-order derivative 
$$\underbrace{
  \frac{1}{H}\frac{d}{dt}\left[\frac{1}{H}\frac{d}{dt}\left[\dots\left[\frac{1}{H}\frac{d}{dt}\right.\right.\right.
}_{n-1}
\left.\left.\left. \left(\frac{dH}{dt}\right) \right]\dots\right]\right]=\frac{1}{2}\frac{d^n E}{dx^n},$$
which also scales as \(L^{-2}\), and we can construct a more general holographic density
\be\label{eq2.2}
\rho_{\cal{H}}=3M_p^2\left[\alpha_0 E+\alpha_1\frac{dE}{dx}+\alpha_2\frac{d^2E}{dx^2}+...\alpha_n\frac{d^n E}{dx^n}\right]
\ee
However, increasing the order of the differential equation imply additional characteristic modes, integration constants and free parameters that potentially lead to problems with initial conditions and parameter degeneracies. We focus on third-order extension that, as shown bellow, allows an exact separation of matter an radiation from the holographic component, a late time accelerating mode and two characteristic modes that scale between matter and radiation, among other configurations. \\

\noindent {\bf The general solution}\\

\noindent The homogeneous solution $E_h$ of the Eq. (\ref{eq2.3}) takes the form
\be\label{eq2.5}
E_h(x)=H_0^2\left(C_1 e^{\lambda_1 x}+C_q e^{\lambda_2 x}+C_3 e^{\lambda_3 x}\right).
\ee
where $\lambda_1,\; \lambda_2,\; \lambda_3$ are the roots of the characteristic equation
\be\label{eq2.4}
\eta\lambda^3+\gamma\lambda^2+\beta\lambda+2(1-\alpha)=0.
\ee
The particular solution $E_p$ is of the form
\be\label{eq2.6}
E_p(x)=H_0^2\left(A_m\Omega_{m0}e^{-3x}+A_r\Omega_{r0}e^{-4x}\right),
\ee
where
\be\label{eq2.7}
A_m=\frac{2}{27\eta-9\gamma+3\beta+2(1-\alpha)},\;\; A_r=\frac{2}{64\eta-16\gamma+4\beta+2(1-\alpha)}
\ee
The complete solution is therefore
\be\label{eq2.8}
E(x)=H^2(x)=H_0^2\left(A_m\Omega_{m0}e^{-3x}+A_r\Omega_{r0}e^{-4x}+\sum_{i=1}^3C_ie^{\lambda_i x}\right)
\ee
In order to avoid residual matter and radiation terms in the holographic component, we impose the restrictions
\be\label{eq2.9}
A_m=1,\;\;\; A_r=1.
\ee
These two conditions define two of the parameters
\be\label{eq2.10}
\alpha=6\gamma-42\eta
\ee
\be\label{eq2.11}
\beta=7\gamma-37\eta
\ee
Thus, in the end, the model is left with two free parameters. Taking into account (\ref{eq2.10}) and  (\ref{eq2.11}) the general solution becomes
\be\label{eq2.12}
H^2(x)=H_0^2\left(\Omega_{m0}e^{-3x}+\Omega_{r0}e^{-4x}+C_1 e^{\lambda_1 x}+C_2 e^{\lambda_2 x}+C_3 e^{\lambda_3 x}\right),
\ee
where $\lambda_1,\; \lambda_2,\; \lambda_3$ are the roots of
\be\label{eq2.13}
\eta\lambda^3+\gamma\lambda^2+\left(7\gamma-37\eta\right)\lambda-2\left(1-6\gamma+42\eta\right)=0
\ee
Note that this solution is free of residual terms that scale as $a^{-3}$ and $a^{-4}$ and were present in the model \cite{Granda2008}. The solution (\ref{eq2.12}) shows a clear separation between matter and holographic components. After the cancelation of the residual matter and radiation components, the entire holographic sector is controlled by two parameters $\gamma$ and $\eta$.
The initial condition $H(0)=H_0$ leads to 
\be\label{eq2.14}
C_1+C_2+C_3=1-\Omega_{m0}-\Omega_{r0}
\ee
Two additional conditions are still required to determine the three constants. These may come, for example, from matching the current value of the equation of state parameter $w_0$ and the deceleration parameter $q_0$, or to require that certain modes not be present for physical reasons like restrictions from early  Universe. To analyze these solutions, let us first begin with the case $\eta = 0$.\\
\subsection*{The case $\eta=0$}
Limiting the model to the second-order equation, making $\eta=0$, the constraints $A_m=1$ and $A_r=1$ reduce to
\be\label{eq2.15}
\alpha=6\gamma,\;\;\; \beta=7\gamma.
\ee
And the Friedmann's  general solution becomes 
\be\label{eq2.16}
H^2(x)=H_0^2=\left(\Omega_{m0} e^{-3x}+\Omega_{r0} e^{-4x}+C_{+}e^{s_{+}x}+C_{-}e^{s_{-}x}\right),
\ee
where
\be\label{eq2.17}
s_{\pm}=\frac{-7\pm \sqrt{1+\frac{8}{\gamma}}}{2}
\ee
Then, the holographic sector becomes controlled by the parameter $\gamma$:
\be\label{eq2.18}
H^2=H_0^2\left(\Omega_{m0} a^{-3}+\Omega_{r0} a^{-4}+C_{+} a^{(-7+\sqrt{1+8/\gamma})/2}+C_{-} a^{(-7-\sqrt{1+8/\gamma})/2}\right).
\ee
In this solution we can distinguish three regimes.\\
\noindent {\bf Region $0<\gamma<1/6$}\\
In this case $\sqrt{1+\frac{8}{\gamma}}>7$, leading to $s_{+}>0$ and $s_{-}<-7$. The  $s_{-}$ mode is clearly problematic since it would inevitably dominate over radiation into the past, affecting BBN and recombination. Therefore, we impose the restriction $C_{-}=0$. The physically relevant solution in then 
\be\label{eq2.19}
H^2=H_0^2\left(\Omega_{m0} a^{-3}+\Omega_{r0} a^{-4}+\Omega_{+} a^{(-7+\sqrt{1+8/\gamma})/2}\right),
\ee
where $\Omega_{+}=C_{+}$. The equation of state parameter for the holographic component is then given by
\be\label{eq2.20}
w_{\cal{H}}=\frac{1}{6}-\frac{1}{6}\sqrt{1+\frac{8}{\gamma}}.
\ee
Since $s_{+}>0$, the equation of state becomes $w_{\cal{H}}<-1$ corresponding to an effective holographic phantom sector.
The $\Lambda$CDM limit is reached for $\gamma=1/6$, for which $s_{+}=0$ and $w_{\cal{H}}=-1$.\\
\noindent {\bf Region $\gamma>1/6$}.\\
In this region
$$1<\sqrt{1+8/\gamma}<7,$$
And therefore 
$$-3<s_{+}<0,\;\; \rm and\;\; -7<s_{-}<-4$$
Again, the $s_{-}$-mode must be removed since it increases faster than radiation towards de past, and the acceptable solution is once again Eq. (\ref{eq2.19}). Since $-3<s_{+}<0$, then the EOS parameter now satisfies $-1<w_{\cal{H}}<0$, corresponding to quintessence behavior. Particularly, as stated before, $\gamma=1/6$ leads to the $\Lambda$CDM, and $\gamma=1$ sets the limit between decelerated and accelerated phases. The solutions in above regions satisfactorily describe the late time cosmic acceleration but cannot contribute to reduce the $H_0$ tension.\\

\noindent{\bf Region $\gamma\le -8$}\\
The modes $s_{\pm}$ satisfy
\be\label{eq2.20a}
-\frac{7}{2}\le s_{+}<-3, \;\;\; -4<s_{-}<-\frac{7}{2}.
\ee
The EOS of the holographic sector in this case is
\be\label{eq2.20b}
w_{\cal{H}}=-1-\frac{1}{3}\frac{s_{+}\Omega_{+}e^{s_{+}x}+s_{-}\Omega_{-}e^{s_{-}x}}{\Omega_{+}e^{s_{+}x}+\Omega_{-}e^{s_{-}x}}
\ee
In the asymptotic case where $s_{+}$-mode dominates, $0<w_{\cal{H}}\le\frac{1}{6}$ and in the case of expansion dominated by the $s_{-}$-mode we find $\frac{1}{6}<w_{\cal{H}}\le\frac{1}{3}$. Then, the model in this region is not viable as dark energy since it can not produce late time accelerated expansion.\\
As will be shown bellow, the inclusion of a third derivative term enriches the cosmological scenarios of late universe and allows us to address the $H_0$ tension problem from a purely geometric point of view, without introducing additional terms of early dark energy.\\
\subsection*{The case $\eta\ne 0$}
In the characteristic equation (\ref{eq2.13}) we can impose a desired value on one of the roots $\lambda$ by fixing one of the free parameters. Let $\lambda=\lambda_{*}$ be the desired value. Rewriting Eq. (\ref{eq2.13}) in the form
\be\label{eq2.21}
(\lambda_{*}+3)(\lambda_{*}+4)\left[\gamma+\eta(\lambda_{*}-7)\right]=2,
\ee
we find
\be\label{eq2.22}
\gamma=\frac{2}{(\lambda_{*}+3)(\lambda_{*}+4)}-\eta(\lambda_{*}-7).
\ee
An important solution, corresponding to a constant mode, is obtained by assuming $\lambda_{*}=0$, which gives the constant term $C_{*}=\Omega_{\Lambda}$ in the solution (\ref{eq2.12}) ($C_1 e^{\lambda_1 x}=C_{*}e^{\lambda_{*} x}=C_{*}=\Omega_{\Lambda}$). For $\lambda_{*}=0$, the conditions (\ref{eq2.22}), (\ref{eq2.10}) and (\ref{eq2.11}) give
\be\label{eq2.23}
\gamma=\frac{1}{6}+7\eta,\;\;\; \alpha=1,\;\;\; \beta=\frac{7}{6}+12\eta
\ee
So far we have a good late-time cosmological solution and a free parameter that allows us to explore the possible cosmological scenarios. To do this, let's analyze the other two roots. With the condition (\ref{eq2.23}), the characteristic equation (\ref{eq2.13}) becomes
\be\label{eq2.25}
\lambda\left[\eta \lambda^2+\left(\frac{1}{6}+7\eta\right)\lambda+\left(\frac{7}{6}+12\eta\right)\right]=0.
\ee
$\lambda=0$ is the root already studied and the other two roots, that depend on $\eta$ are
\be\label{eq2.26}
\lambda_{2,3}=\frac{-\left(\frac{1}{6}+7\eta\right)\pm \sqrt{\eta^2-\frac{7}{3}\eta+\frac{1}{36}}}{2\eta}
\ee
The roots are real for
\be\label{eq2.28}
\eta\le\eta_{-}\;\;\; \rm or \;\;\; \eta\ge\eta_{+}.
\ee
where $\eta_{\pm}$ are the zeros of the discriminant
$$\eta_{\pm}=\frac{7\pm 4\sqrt{3}}{6}.$$
In the interval
$$\eta_{-}< \eta< \eta_{+},$$ 
the roots form a complex conjugate pair.
The product and the sum of the roots are
\be\label{eq2.29}
\lambda_2+\lambda_3=-7-\frac{1}{6\eta};\;\;  \lambda_2\lambda_3=12+\frac{7}{6\eta}
\ee
This helps us to analyze the different regions for the solutions, that are shown in Tables I and II.\\
\noindent {\bf Modes for $\eta>0$}\\
\begin{center}
\begin{tabular}{lccc}\hline\hline
 Region  \hspace{0.7cm} & $\lambda_2$  \hspace{0.7cm}  & $\lambda_3$\\ 
\hline
I: $\eta<\eta_{-}$ & $-7-2\sqrt{3}<\lambda_2<-7$ & $\lambda_3<-7-2\sqrt{3}$ \\
II: $\eta>\eta_{+}$ & $-7+2\sqrt{3}<\lambda_2<-3$ & $-4<\lambda_3<-7+2\sqrt{3}$ \\ 
 \hline
\end{tabular}
\end{center}
\begin{center}
{\bf Table I: Modes for  $\eta>0$}
\end{center}

\begin{center}
{\it The region I is not physical since both modes increase faster than radiation towards the past, affecting the BBN. In region II both modes scale between matter and radiation, where one mode is closer to matter while the other is closer to radiation. Both modes in region II could be of physical interest, provided their amplitudes are small enough not to conflict with the early-Universe restrictions.}
\end{center}

\noindent {\bf Modes for $\eta<0$}\\
\begin{center}
\begin{tabular}{lccc}\hline\hline
 Region  \hspace{0.7cm} & $\lambda_2$  \hspace{0.7cm}  & $\lambda_3$\\ 
\hline
I: $-\frac{1}{42}<\eta<0$ & $-7<\lambda_2<-\sqrt{37}$ & $\lambda_3>\sqrt{37}$ \\
II: $-\frac{7}{72}<\eta<-\frac{1}{42}$ & $-\sqrt{37}<\lambda_2<-\frac{37}{7}$ & $0<\lambda_3<\sqrt{37}$ \\ 
III: $\eta<-\frac{7}{72}$ & $-\frac{37}{7}<\lambda_2<-4$ & $-3<\lambda_3<0$ \\ 
 \hline
\end{tabular}
\end{center}
\begin{center}
{\bf Table II: Modes for  $\eta<0$}
\end{center}
\begin{center}
{\it The region I has no physical solutions since the $\lambda_2$-mode increases faster than radiation towards the past, and the $\lambda_3$-mode is too phantom-like. In region II the $\lambda_2$-mode is excluded for the same reasons as before, and the mode $\lambda_3$ gives rise to a phantom expansion. In region III the mode $\lambda_2$ is excluded and the mode $\lambda_3$ gives rise to quintessential expansion.}
\end{center}
The mode $\lambda_3$ gives rise to a phantom sector which, added to the cosmological constant ($\lambda_1$-mode) produces an expansion with an equation of state below the cosmological constant limit. From all of the above regions we can 
 extract three different cosmological scenarios.\\

\noindent{\bf I.} The first scenario is the solution for $\eta>0$ in the region $\eta>\eta_{+}$, where the cosmological expansion is  described by 
\be\label{eq2.30}
H^2(x)=H_0^2\left[\Omega_{m0} e^{-3x}+\Omega_{r0} e^{-4x}+\Omega_{0}+\Omega_2e^{\lambda_2 x}+\Omega_3e^{\lambda_3 x}\right],
\ee
where $\Omega_{0}=C_1$ is the amplitude for the zero-mode, $\Omega_i=C_i$ ($i=2,3$) and are subject to restriction
\be\label{eq2.30a}
\Omega_{m0}+\Omega_{r0}+\Omega_{0}+\Omega_2+\Omega_3=1.
\ee
The dark energy sector (holographic sector) of this solution leads to the following EOS
\be\label{eq2.31}
w_{\cal{H}}=-1-\frac{1}{3}\frac{ \lambda_2\Omega_2e^{\lambda_2 x}+ \lambda_3\Omega_3e^{\lambda_3 x}}{\Omega_0+\Omega_2e^{\lambda_2 x}+\Omega_3e^{\lambda_3 x}},
\ee
where $-7+2\sqrt{3}<\lambda_2<-3$ and $-4<\lambda_3<-7+2\sqrt{3}$. Since $\lambda_2,\lambda_3<0$, then $w_{\cal{H}}>-1$ and the solution is of the quintessence type. The correction to the EOS of the cosmological constant ($w_{\Lambda}=-1$)  depends on the fraction
$$\frac{\lambda_2\Omega_2 e^{\lambda_2 x}+\lambda_3\Omega_3 e^{\lambda_3 x}}{\Omega_{0}+\Omega_2e^{\lambda_2 x}+\Omega_3e^{\lambda_3 x}}$$
which should be small according to observations. The effective EOS is 
\be\label{eq2.32}
w_{eff}=-1+\frac{3\Omega_{m0} e^{-3x}+4\Omega_{r0} e^{-4x}-\lambda_2\Omega_2e^{\lambda_2 x}-\lambda_3\Omega_3e^{\lambda_3 x}}{3\left[\Omega_{m0} e^{-3x}+\Omega_{r0} e^{-4x}+\Omega_{0}+\Omega_2e^{\lambda_2 x}+\Omega_3e^{\lambda_3 x}\right]},
\ee
Given the intervals satisfied by $\lambda_2$ and $\lambda_3$, we can note that before matter dominance there is a period, between the end of radiation dominance and the beginning matter dominance, when the terms $\Omega_2e^{\lambda_2 x}=\Omega_2 a^{\lambda_2}$ and $\Omega_3e^{\lambda_3 x}=\Omega_3 a^{\lambda_3}$ could be relevant. These terms do not substantially affect late-universe cosmological behavior. However, they can lead to an increase in $H$ during the period between decoupling and matter-radiation equality, and therefore deserve a separate discussion.\\
In Fig.1 we describe the late-time evolution of  $w_{eff}$ compared to $w_{\Lambda CDM}$ and the relative deviation of the Hubble parameter with respect to $H_{\Lambda CDM}$ in the redshift interval $10^3\le z\le 10^4$.
For this numerical case we consider the modes $\lambda_1=-17/5$ and $\lambda_2=-11/3$ and take $\Omega_{m0}=0.315$,  $\Omega_{r0}=9.2\times 10^{-5}$, $\Omega_2=10^{-3}$ and $\Omega_3=10^{-4}$.
\begin{figure}
\begin{center}
\includegraphics[scale=0.58]{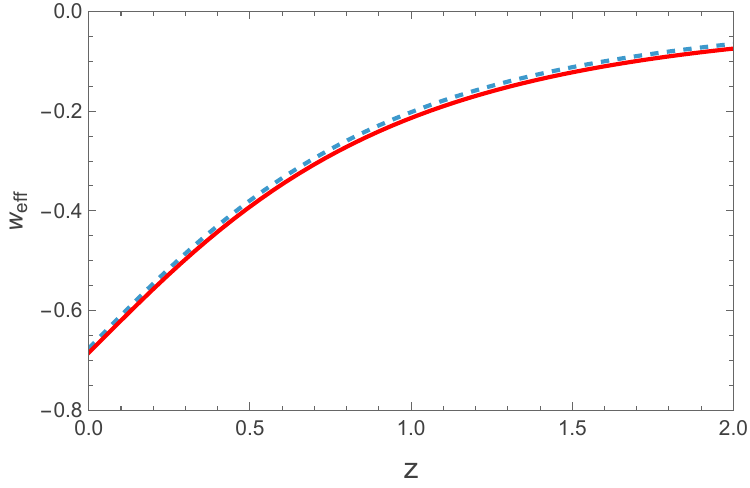}
\includegraphics[scale=0.58]{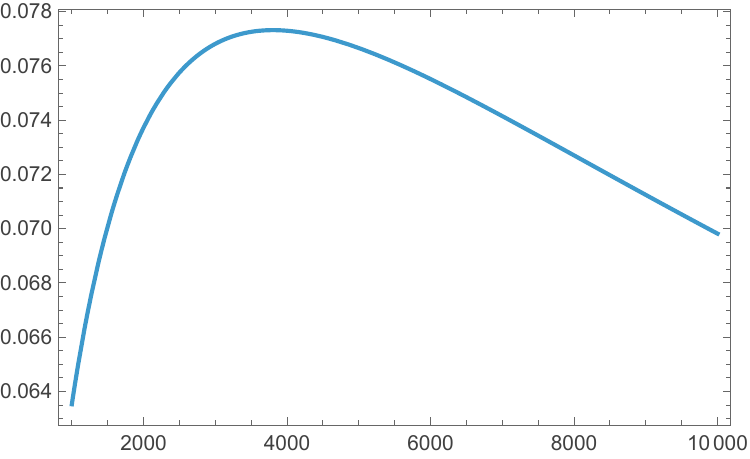}
\caption{\it The late-time evolution of $w_{eff}(z)$ ($0<z<2$) for the model(\ref{eq2.31}) compared to $w_{\Lambda CDM}$ (red). The right plot shows the relative deviation of the Hubble parameter with respect to $H_{\Lambda CDM}$ in the region before and after the matter-radiation equality ($10^3<z<10^4$). Note the increment of $H$ with respect to $H_{\Lambda}$ in the region around the matter-radiation equilibrium. At the time of BBN ($z\sim 10^9$) the relative deviation ($\Delta H/H_{\Lambda CDM})\sim 10^{-3}$. } 
\label{fig1}
\end{center}
\end{figure}
\noindent {\bf II.} The second scenario is the solution for $\eta<0$ in region $-7/72<\eta<-1/42$, where $0<\lambda_3<\sqrt{37}$ and $C_2=0$.
This mode, combined with the constant density term, results in phantom behavior. The analytical expression for this solution follows from Eqs. (\ref{eq2.30})-(\ref{eq2.32}), by setting  $\Omega_2=0$.
The equation of state of the holographic component, given by
\be\label{eq2.33}
w_{\cal{H}}=-1-\frac{\lambda_3}{3}\frac{\Omega_3e^{\lambda_3 x}}{\Omega_0+\Omega_3e^{\lambda_3 x}},
\ee
takes values below \(-1\) for $0<\lambda_3<\sqrt{37}$. In this solution, the holographic component corresponding to $\lambda_3$-mode is negligible in early times.\\ 
\noindent{\bf III.} The third scenario is the solution for $\eta<-7/72$, where the $\lambda_3$-mode is in the interval $[-3,0]$ and $C_2=0$. In this interval, the EOS for the holographic component is given by (\ref{eq2.33}), but this time $\lambda_3<0$ gives rise to quintessential expansion since $-1<w_{\cal{H}}<0$.\\
\noindent The analytical simplicity of all the above solutions makes them easy to adapt to observational constraints. 
\section{Sound Horizon Reduction and the $H_0$ Tension}
In this section we study whether the homogeneous modes generated by the extended local infrared cutoff can modify the pre-recombination expansion history enough to reduce the comoving sound horizon, while remaining compatible with early-Universe constraints. We analyze the first scenario corresponding to the region $\eta>\eta_{+}$ in table I.\\
The expansion rate can be written as
\be\label{eq3.1}
\tilde{E}(z)=\frac{H^2(z)}{H_0^2}=\Omega_{r0}(1+z)^4+\Omega_{m0}(1+z)^3+\Omega_{0}+\Omega_2(1+z)^{-\lambda_2}+\Omega_3(1+z)^{-\lambda_3},
\ee
The normalization condition ($H^2/H_0^2=1$) fixes the amplitude of the constant mode,
\be\label{eq3.2}
\Omega_0=1-\Omega_{m0}-\Omega_{r0}-\Omega_2-\Omega_3.
\ee
We take the $\Lambda$CDM background as reference, with the same present matter and
radiation densities
\be\label{eq3.3}
\tilde{E}_{\Lambda{\rm CDM}}(z)=\Omega_{r0}(1+z)^4+\Omega_{m0}(1+z)^3+\Omega_{\Lambda},
\ee
where $$\Omega_{\Lambda}=1-\Omega_{m0}-\Omega_{r0},$$
so that the modification introduced by the holographic modes is exactly
\be\label{eq3.4}
\Delta \tilde{E}(z)\equiv \tilde{E}(z)-\tilde{E}_{\Lambda{\rm CDM}}(z)=
\Omega_2\left[(1+z)^{-\lambda_2}-1\right]+\Omega_3\left[(1+z)^{-\lambda_3}-1\right].
\ee
Given that $\Omega_2, \Omega_3>0$, both modes increase the expansion rate at $z>0$. Their effect, however, depends strongly on redshift, since the two modes have different scaling properties between the matter and radiation ones.\\
\noindent The EOS of the holographic sector in (\ref{eq3.1}) is  
\be\label{eq3.4a}
w_{\cal{H}}=-1-\frac{1}{3}\frac{\lambda_2\Omega_2e^{\lambda_2 x}+\lambda_3\Omega_3e^{\lambda_3 x}}{\Omega_0+\Omega_2e^{\lambda_2 x}+\Omega_3e^{\lambda_3 x}},
\ee
and since $\lambda_2,\lambda_3<0$ it satisfies $w_{\cal{H}}>-1$, that is, the holographic
sector is of quintessence type.\\
\subsection* {The Sound Horizon}

Now let us analyze the effect of $\lambda_2,\; \lambda_3$ modes in the evaluation of the sound horizon. The sound horizon at the baryon drag epoch is
\be\label{eq3.8}
r_s(z_d)=\int_{z_d}^{\infty}\frac{c_s(z)}{H(z)}\,dz,
\ee
where the sound speed of the tightly coupled photon-baryon fluid is
\be\label{eq3.9}
c_s(z)=\frac{c}{\sqrt{3[1+R(z)]}}, \qquad
R(z)=\frac{3\omega_b}{4\omega_\gamma}\,\frac{1}{1+z},
\ee
and $\omega_b=\Omega_{b0}h^2$, $\omega_\gamma=\Omega_{\gamma0}h^2$
are the physical baryon and photon densities. Then,
\be\label{eq3.11}
r_s(z_d)=\frac{c}{H_0\sqrt{3}}\int_{z_d}^{\infty}\frac{dz}
{\sqrt{\left[1+\frac{3\omega_b}{4\omega_\gamma}\frac{1}{1+z}\right]\tilde{E}(z)}}.
\ee
Equation (\ref{eq3.11}) shows the mechanism by which the homogeneous modes modify the acoustic scale: $\tilde{E}(z)$ increases in the integration interval and, therefore decreases $r_s$. Using (\ref{eq3.3}) the sound horizon takes the form
{\small
\be\label{eq3.12}
r_s(z_d)=\frac{c}{H_0\sqrt{3}}\int_{z_d}^{\infty}\frac{dz}{\sqrt{1+R(z)}
\sqrt{\tilde{E}_{\Lambda{\rm CDM}}(z)+\Omega_2\left[(1+z)^{-\lambda_2}-1\right]+\Omega_3\left[(1+z)^{-\lambda_3}-1\right]}}.
\ee}
which makes explicit the positive correction with respect to $\Lambda$CDM and the corresponding reduction of $r_s$ with respect to $r_s^{\Lambda{\rm CDM}}$. \\
The sound horizon integral can also be written in terms of the physical densities $\omega_b$, $\omega_{\gamma}$, $\omega_m$ and $\omega_r$ as follows from (\ref{eq3.1}) 
\small{
\be\label{eq3.13}
r_s(z_d)=\frac{c}{100\sqrt{3}}\int_{z_d}^{\infty}\frac{dz}{\sqrt{1+R(z)}
\sqrt{\omega_m(1+z)^3+\omega_r(1+z)^4+\omega_0+\omega_2(1+z)^{-\lambda_2}+\omega_3(1+z)^{-\lambda_3}}},
\ee}
where $H_0^2=(100h)^2$ and we defined, by analogy with physical densities, 
$$\omega_0=h^2\Omega_0,\;\; \omega_2=h^2\Omega_2,\;\; \omega_3=h^2\Omega_3.$$ 
With 
\be\label{eq3.14}
\frac{H^2}{100^2}=\omega_m(1+z)^3+\omega_r(1+z)^4+\omega_0+\omega_2(1+z)^{-\lambda_2}+\omega_3(1+z)^{-\lambda_3}
\ee
The normalization condition now turns into
$$\omega_0=h^2-\omega_m-\omega_r-\omega_2-\omega_3.$$
For the physical densities we use the values
\be
\omega_b=0.02237,\quad \omega_\gamma=2.469\times10^{-5},\quad
\omega_m=0.143097,\quad \omega_r=4.1793\times10^{-5},
\ee
while $\omega_2$, $\omega_3$ are constrained mainly by BBN and by matter-radiation equality.
Given the ranges of $\lambda_2$ and $\lambda_3$, in the integration  interval $z\ge 1059$ the term $\omega_0$ is completely negligible and can be omitted from (\ref{eq3.13}), in the same way as $\Omega_0$ from (\ref{eq3.12}).
Then, the sound horizon is independent of $h$ and the integrals (\ref{eq3.12}) and (\ref{eq3.13}) give the same result. 
At late times, however, there is an additional mechanism that can further increase $H_0$. This mechanism is analyzed in the next section.

\subsection*{Reduction of the Sound Horizon with One and Two modes}
\noindent The densities associated with the homogeneous modes scale as
\be\label{eq3.5}
\rho_2\propto(1+z)^{-\lambda_2},
\qquad
\rho_3\propto(1+z)^{-\lambda_3}.
\end{equation}
Both modes grow towards the past faster than matter but slower than radiation. In
particular, $\rho_2/\rho_r\propto(1+z)^{-\lambda_2-4}$ and $\rho_3/\rho_r\propto(1+z)^{-\lambda_3-4}$. Since $\lambda_2,\lambda_3>-4$, both ratios decrease at sufficiently high redshift. In consequence, the holographic correction can be appreciable around recombination and matter-radiation equality, and becomes progressively less important towards BBN.\\
We define the fractional contribution of the two dynamical holographic modes as
\be\label{eq3.7}
f_{\cal{H}}(z)=\frac{\Omega_2(1+z)^{-\lambda_2}+\Omega_3(1+z)^{-\lambda_3}}{\tilde{E}(z)},
\end{equation}
which can be used as a criterion to control the size of the holographic component at early times, in particular at matter-radiation equality.\\

\noindent We consider first the effect of the $\lambda_2$ mode alone, which is closer to the matter scaling than the $\lambda_3$ mode. This mode concentrates the additional contribution around $z\sim z_{eq}$ and becomes irrelevant rapidly towards BBN. For this first analysis we set $\Omega_3=0$ ($\omega_3=0$) in Eqs. (\ref{eq3.1})-(\ref{eq3.14}). 
The fractional contribution of the $\lambda_2$ mode, as follows from (\ref{eq3.7}), is given by
\be\label{eq3.15}
f^{(2)}_{\cal{H}}(z)=\frac{\Omega_2(1+z)^{-\lambda_2}}{\Omega_{m0}(1+z)^3+\Omega_{r0}(1+z)^4+\Omega_2(1+z)^{-\lambda_2}},
\ee
where we neglected $\Omega_0$ for $z\ge z_d$.\\
In this study we take into account the present limits on early dark energy (EDE) set by different observations. Planck NPIPE data \cite{PlanckNPIPE} give approximately $f_{EDE}<0.061$ at $95\%$ c.l. in the interval $10^3\lesssim z\lesssim 10^4$, while Planck+CMB lensing+DESI combinations \cite{DESI2025} allow values of order $0.09$. When Ly-$\alpha$ data are included \cite{DESI2025a}, the bounds on EDE can be lowered to $f_{EDE}<0.03$. It is therefore useful to have three reference levels: approximately $3\%$, $5\%$ and $10\%$. \\
As a representative value we take $\lambda_2=-10/3$, with 
\be
\Omega_{m0}=0.315,\quad \Omega_{r0}=9.2\times10^{-5},\quad z_d=1059,\quad z_{eq}=3400,
\ee
and fix $\Omega_2$ from Eq. (\ref{eq2.15}) for each value of $f^{(2)}_{\cal{H}}(z_{eq})$. The results are shown in table I.
These results were obtained within the usual approximation in which the acoustic calibration keeps $r_sH_0$ approximately constant, with the reference values
$H_0^{\Lambda{\rm CDM}}=67.4$~km\,s$^{-1}$\,Mpc$^{-1}$ and
$r_s^{\Lambda{\rm CDM}}=147.01$~Mpc.

\begin{center}
\begin{tabular}{lccccccc}\hline\hline
 $f^{(2)}_{\cal{H}}(z_{eq})$  \hspace{0.4cm} & $\Omega_2$  \hspace{0.4cm}  &$f^{(2)}_{\cal{H}}(z_{d})$  \hspace{0.4cm} & $f^{(2)}_{\cal{H}}(z_{BBN})$ \hspace{0.4cm} & $r_s [Mpc]$ \hspace{0.4cm} &$\Delta r_s/r_s$\  \hspace{0.4cm} &$H^{\cal{H}}_0$\\ 
\hline
$3\%$ & $1.291\times 10^{-3}$ & $3.093\%$ & $0.0014\%$ & $144.987$ & $-1.376\%$ & $68.341$\\ 
$5\%$& $2.198\times 10^{-3}$  &$5.152\%$ &$0.0024\%$ &$143.616$ & $-2.309\%$ & $68.993$\\
$10\%$& $4.639\times 10^{-3}$  &$10.287\%$ &$0.0050\%$ &$140.119$ & $-4.687\%$ & $70.715$\\
 \hline
\end{tabular}
\end{center}
\begin{center}
{\bf Table I}
\end{center}
For the most conservative case, $f^{(2)}_{\cal{H}}(z_{eq})=3\%$, we obtain $\Delta r_s/r_s\simeq -1.38\%$ and $H^{\cal{H}}_0\simeq 68.34 \; \rm{km \;s^{-1}Mpc^{-1}}$, while for the extreme value  $f^{(2)}_{\cal{H}}(z_{eq})=10\%$ we find  $\Delta r_s/r_s\simeq -4.69\%$ and $H^{\cal{H}}_0\simeq 70.72 \; \rm{km \;s^{-1}Mpc^{-1}}$. 
Thus, in order to reach $H_0\simeq 70.7$ it necessary to use almost completely a contribution of order $f^{(2)}_{\cal{H}}(z_{eq})\simeq 10\%$ around equality. The observational question is then how much of that $5\%$-$10\%$ around $z$-equality, can be tolerated by CMB observations once the perturbations are taken into account.\\
The same analysis for $\lambda_2=-7/2$, which is halfway between $-3$ and $-4$, is summarized in table II.
\begin{center}
\begin{tabular}{lccccccc}\hline\hline
 $f^{(2)}_{\cal{H}}(z_{eq})$  \hspace{0.4cm} & $\Omega_2$  \hspace{0.4cm}  &$f^{(2)}_{\cal{H}}(z_{d})$  \hspace{0.4cm} & $f^{(2)}_{\cal{H}}(z_{BBN})$ \hspace{0.4cm} & $r_s [Mpc]$ \hspace{0.4cm} &$\Delta r_s/r_s$\  \hspace{0.4cm} &$H^{\cal{H}}_0$\\ 
\hline
$3\%$ & $3.330\times 10^{-4}$ & $2.561\%$ & $0.0114\%$ & $145.018$ & $-1.355\%$ & $68.326$\\ 
$5\%$& $5.667\times 10^{-4}$  &$4.281\%$ &$0.0195\%$ &$143.667$ & $-2.274\%$ & $68.968$\\
$10\%$& $1.1963\times 10^{-3}$  &$8.627\%$ &$0.0411\%$ &$140.211$ & $-4.625\%$ & $70.668$\\
 \hline
\end{tabular}
\end{center}
\begin{center}
{\bf Table II}
\end{center}
Compared to $\lambda_2=-10/3$ the difference is minimal. The value $\lambda_2=-10/3$ gives a slightly greater reduction of $r_s$ for the same fraction, while for $\lambda_2=-3.5$ the present amplitude is much smaller and its late background deviation is also small. In both cases, BBN remains completely safe.\\

\noindent Turning to the two-mode solution (\ref{eq3.1}) we now evaluate it's effect on the reduction of the sound horizon.
The ratio between the densities of the two modes evolves as
\be\label{eq3.16}
\frac{\rho_3}{\rho_2}=\frac{\Omega_3}{\Omega_2}(1+z)^{\lambda_2-\lambda_3}=\frac{\Omega_3}{\Omega_2}(1+z)^{\delta},
\ee
where $\delta=\lambda_2-\lambda_3>0$, as follows from $\lambda_2$,$\lambda_3$ intervals. This ratio grows towards the past because the $\lambda_3$ mode grows faster than the $\lambda_2$ mode. For this reason the near-matter mode gives the larger modification precisely in the interval $z_d\lesssim z\lesssim  z_{eq}$, where it is more efficient in reducing $r_s$. \\
\noindent Considering the extreme condition $f_{\cal{H}}(z_{eq})=10\%$, in table III we show the numerical results for different percentages of the $\lambda_3$ mode. In this case the modes take the values $\lambda_2=-10/3$ and $\lambda_3=-41/11$ (corresponding to $\eta=11/4$). The condition $f_{\cal{H}}(z_{eq})=10\%$ gives a linear relation between $\Omega_2$ and $\Omega_3$. Defining 
\be\label{eq3.17}
q_3=\frac{\rho_3(z_{eq})}{\rho_2(z_{eq})+\rho_3(z_{eq})},
\ee
we obtain the results of table III.
\begin{center}
\begin{tabular}{lccccccc}\hline\hline
 $q_3 (\%)$  \hspace{0.4cm} & $\Omega_2$  \hspace{0.4cm}  & $\Omega_3$  \hspace{0.4cm} & $f_{\cal{H}}(z_{d})$  \hspace{0.4cm} & $f_{\cal{H}}(z_{BBN})$ \hspace{0.4cm} & $\Delta r_s/r_s$\  \hspace{0.4cm} &$H^{\cal{H}}_0$\\ 
\hline
$0\%$ & $4.639\times 10^{-3}$ & $0$ & $10.287\%$ & $0.005\%$ & $-4.687\%$ & $70.715$\\ 
$25\%$& $3.479\times 10^{-3}$  &$4.711\times 10^{-5}$ &$9.429\%$ &$0.183\%$ & $-4.695\%$ & $70.721$\\
$50\%$& $2.320\times 10^{-3}$  &$9.423\times 10^{-5}$ &$8.555\%$ &$0.361\%$ & $-4.699\%$ & $70.723$\\
$75\%$& $1.160\times 10^{-3}$  &$1.413\times 10^{-4}$ &$7.664\%$ &$0.538\%$ & $-4.698\%$ & $70.723$\\
$100\%$& $0$  &$1.885\times 10^{-4}$ &$6.755\%$ &$0.714\%$ & $-4.693\%$ & $70.719$\\
 \hline
\end{tabular}
\end{center}
\begin{center}
{\bf Table III}
\end{center}
Note that all cases of table III satisfy $f_{\cal{H}}(z_{BBN})<1\%$. Therefore, BBN does not impose a strong restriction on these two modes.
It is also worth noting that, for the pair $(-10/3,-41/11)$ the sound horizon depends very weakly on the mixture, since the
resulting values of $H_0$ are practically the same.\\
The difference appears mainly at low redshift. For $q_3=50\%$ the deviation of the Hubble parameter at $z=2$ is 
\be\label{eq3.18}
\frac{\Delta H}{H^{\Lambda CDM}_0}(z=2)\simeq 0.507\%,
\ee
while for the pure $\lambda_2=-10/3$ case,
\be\label{eq3.19}
\frac{\Delta H}{H^{\Lambda CDM}_0}(z=2)\simeq 0.952\%,
\ee
So including a fraction of the $\lambda_3=-41/11$ mode improves the late-time background without practically affecting the reduction of the acoustic horizon.\\

\noindent {\bf Effective matter-radiation equality}\\
\noindent An important indicator is the shift of the effective matter-radiation equality. The fundamental equality for 
$\rho_m(z_{eq})=\rho_r(z_{eq})$ remains unchanged in this model as the residual terms were exactly eliminated. What changes is the moment at which the total expansion behaves as an equal mixture of matter and radiation. We
define this effective equality by the condition
\be\label{eq3.17a}
w_{eff}(z_{eq}^{eff})=\frac{1}{6}
\ee
which is the moment when the total equation of state takes on the intermediate value between matter ($w=0$) and radiation ($w=1/3$)
This is the same value that takes place in the fundamental equality between matter and radiation when $\rho_m=\rho_r$:
\be
w(z_{eq})=\frac{\frac{1}{3}\rho_r(z_{eq})}{\rho_m(z_{eq})+\rho_r(z_{eq})}=\frac{\frac{1}{3}\rho_r}{2\rho_r}=\frac{1}{6}.
\ee
With the equations of state of the two modes, 
\be
w_2=-1-\frac{\lambda_2}{3},\qquad w_3=-1-\frac{\lambda_3}{3},
\ee
and neglecting the constant mode $\Omega_0$, which is irrelevant at these redshifts, the
effective equation of state is
\be\label{eq3.17b}
w_{eff}=\frac{\frac{1}{3}\rho_r+w_2\rho_2+w_3\rho_3}{\rho_m+\rho_r+\rho_2+\rho_3}
\ee
Imposing the condition (\ref{eq3.17a}) gives
\be\label{eq3.17c}
\rho_m-\rho_r+(2\lambda_2+7)\rho_2+(2\lambda_3+7)\rho_3=0
\ee
Since $\lambda_2>-7/2$ and $\lambda_3<-7/2$, we have $2\lambda_2+7>0$ and $2\lambda_3+7<0$,
so the two modes shift $z_{eq}^{\rm eff}$ in opposite directions. This allows to choose the
amplitudes in such a way that
\be\label{eq3.17d}
(2\lambda_2+7)\rho_2(z_{eq})+(2\lambda_3+7)\rho_3(z_{eq})\simeq 0
\ee
For the pair $\lambda_2=-10/3$, $\lambda_3=-41/11$, the condition (\ref{eq3.17d}) gives $q_3=11/26$. The results for this value of $q_3$ are shown in Table IV.
\begin{center}
\begin{tabular}{lccccccc}\hline\hline
 $f_{\cal{H}}(z_{eq})$  \hspace{0.1cm} & $\Omega_2$  \hspace{0.3cm}  & $\Omega_3$  \hspace{0.3cm} & $f_{\cal{H}}(z_d)$  \hspace{0.3cm}& $f_{\cal{H}}(z_{BBN})$  \hspace{0.2cm} & $r_d\; {\rm [Mpc]}$ \hspace{0.2cm} & $\Delta r_d/r_d$\  \hspace{0.3cm} &$H^{\cal{H}}_0$\\ 
\hline
$3\%$ & $7.450\times 10^{-4}$ & $2.219\times 10^{-5}$ & $2.624\%$ & $0.085\%$ & $144.986$ & $-1.377\%$ & $68.341$\\ 
$5\%$& $1.268\times 10^{-3}$  &$3.777\times 10^{-5}$ & $4.384\%$ & $0.145\%$ &$143.613$ & $-2.311\%$ & $68.994$\\
$10\%$& $2.676\times 10^{-3}$  &$7.973\times 10^{-5}$ & $8.826\%$ &$0.306\%$ &$140.103$ & $-4.698\%$ & $70.723$\\
 \hline
\end{tabular}
\end{center}
\begin{center}
{\bf Table IV}
\end{center}
Compared to Table III, Table IV (for the condition $f_{\cal{H}}(z_{eq})=10\%$) shows that the acoustic horizon can be reduced by practically the same amount, while the shift of the effective matter-radiation equality is cancelled to first order.                         

\section{Phantom Mode and Late Time Amplification of $H_0$}

\noindent In the previous section the root $\lambda_1=0$ was assumed, so that the corresponding
mode behaves as a cosmological constant and, together with the two early modes, leads
to an effective expansion of quintessence type. We now consider a slightly positive root,
\be\label{eq3.18b}
0.15\le\lambda_1\le0.30,
\ee
which corresponds to an effective phantom component with equation of state
\be\label{eq3.19b}
w_1=-1-\frac{\lambda_1}{3},  \qquad -1.10 \le w_1 \le -1.05 .
\ee
From Eqs. (\ref{eq2.21}), (\ref{eq2.22}) and the characteristic equation it follows that the other two roots remain in the interval $(-4,-3)$, but the point separating both branches is slightly displaced. For $\lambda_1=0$ the two roots coincide at $-7+2\sqrt{3}\simeq -3.53590$. The intervals for $\lambda_2$ and $\lambda_3$ corresponding
to some values of $\lambda_1$ are shown in Table V.

\begin{center}
\begin{tabular}{lccc}\hline\hline
 $\lambda_1$  \hspace{0.7cm} & $\lambda_2$  \hspace{0.7cm}  & $\lambda_3$\\ 
\hline
$0$ & $-3.53590<\lambda_2<-3$ & $-4<\lambda_3<-3.53590$ \\
$0.15$ & $-3.53441<\lambda_2<-3$ & $-4<\lambda_3<-3.53441$ \\ 
$0.30$ & $-3.53304<\lambda_2<-3$ & $-4<\lambda_3<-3.53304$ \\
 \hline
\end{tabular}
\end{center}
\begin{center}
{\bf Table V}
\end{center}
Therefore, the shift of $\lambda_1$ from $0$ to $0.15 - 0.30$ has practically no effect on the window of the two early roots
and the early-time results of Tables I-IV remain unchanged. This is due to the fact that for $z\gtrsim z_d$, the term $\Omega_1 (1+z)^{-\lambda_1}$ is negligible compared to matter, radiation, and the $\lambda_2, \lambda_3$ modes. Nevertheless, the phantom mode has a non-trivial effect on the late-time geometry, which
we analyze in what follows.\\
\subsection*{One early mode}
We first consider only the \(\lambda_2\) mode and set $\omega_3=0$. The Hubble parameter then reads\\
\be\label{eq3.19c}
H^2(z)=100^2\left[\omega_m(1+z)^3+\omega_r(1+z)^4+\omega_1(1+z)^{-\lambda_1}+\omega_2(1+z)^{-\lambda_2}\right],
\ee
with
\be\label{eq3.20}
\omega_1=h^2-\omega_m-\omega_r-\omega_2.
\ee
In section 3 the Hubble parameter was estimated assuming that the acoustic calibration
keeps the product $r_s H_0$ approximately constant,
\be\label{eq3.21}
H_0\simeq H^{\Lambda CDM}_0\frac{r_s^{\Lambda}}{r_s}.
\ee
This estimate is no longer adequate when $\lambda_1>0$, since according to(\ref{eq3.20}) a change in $h$ also modifies $\omega_1$. In fact, since $\rho_1=\rho_{1,0}(1+z)^{-\lambda_1}$ with $w_1<-1$, the factor $(1+z)^{-\lambda_1}$ is
smaller than unity for $z>0$. Thus, for the same present value $\omega_1$, the phantom component contributes less than a cosmological constant at $z>0$. This means that it can have a relatively large density today, which increases $H_0$, and decrease rapidly towards $z\sim1$--$2$. In this way a larger value of $H_0$ may be obtained without a significant
increase of $H(z)$ in the redshift range probed by BAO and supernovae.\\
To quantify this effect we keep the angular acoustic scale fixed. The comoving distance to
the last scattering surface is
\be\label{eq3.22}
D_M(z_{*})=\frac{c}{100\sqrt{3}}\int_{0}^{z_{*}}\frac{dz}{
\sqrt{\omega_m(1+z)^3+\omega_r(1+z)^4+\omega_2(1+z)^{-\lambda_2}+(h^2-\omega_m-\omega_r-\omega_2)(1+z)^{-\lambda_1})}},
\ee
and we impose
\be\label{eq3.23}
\theta_s=\frac{r_s(z_{*})}{D_M(z_{*})} = \theta_s^{\Lambda}.
\ee
Eq. (\ref{eq3.23}) can be solved for $h=h(\lambda_1,\lambda_2,\omega_2)$ instead of using the approximation $H_0\propto 1/r_s$ (\ref{eq3.21}).\\
As a benchmark we take the point $\lambda_1=0.15,\;\; \lambda_2=-3.3$, with $f_{\cal{H}}(z_{eq})=5\%$ giving approximately $w_2\simeq 1.32\times 10^{-3}$. 
For the reference $\Lambda$CDM model we have $r^{\Lambda}_s(z_{*})\simeq 144.27\; \rm{Mpc}$ and $D_M^{\Lambda}(z_{*})\simeq 13864.3\; \rm{Mpc}$, so that $100\theta_s^{\Lambda}=1.04060$,  while for the benchmark $r_s(z_{*})=140.93 \rm Mpc$.
Keeping $\theta_s$ fixed then requires
\be\label{eq3.24}
D_M(z_{*})=\frac{r_s(z_{*})}{\theta_s^{\Lambda}}\simeq 13543.15 \;\rm{Mpc},
\ee
and the numerical solution of Eq.~(\ref{eq3.24}) gives
\be\label{eq3.25}
h\simeq0.7472, \qquad H_0\simeq74.72~{\rm km\,s^{-1}\,Mpc^{-1}}.
\ee
This value is considerably larger than the one obtained from Eq.~(\ref{eq3.21}),
$H_0\simeq67.4\,r_s^{\Lambda}/r_s\simeq69.0~{\rm km\,s^{-1}\,Mpc^{-1}}$, which shows that,
at fixed physical densities, the distance to the last scattering surface depends
non-trivially on $h$ when $\lambda_1>0$.
This amplification, however, has a cost in the late-time background. For $\lambda_1=0.15$
and $h=0.7472$ the relative deviations of the Hubble parameter are
$$\frac{\Delta H}{H_{\Lambda}}\Big |_{z=0}=10.86\%, \;\frac{\Delta H}{H_{\Lambda}}\Big |_{z=0.5}=5.12\%, 
\; \frac{\Delta H}{H_{\Lambda}}\Big |_{z=1}=2.54\%, \; \frac{\Delta H}{H_{\Lambda}}\Big |_{z=2}=1.06\%.$$
A deviation of about $5\%$ in $H(z)$ around $z\simeq0.5$ is expected to be in strong
tension with BAO and supernova data, so the condition on $\theta_s$ alone is not sufficient.
The value $H_0\simeq74.7~{\rm km\,s^{-1}\,Mpc^{-1}}$ must therefore be interpreted as the
maximum geometric degeneracy allowed by $\theta_s$.
To determine how much of this amplification remains, we impose simultaneously a BAO
distance around $z\simeq0.5$, through the ratios $D_M(z)/r_d$ and $D_H(z)/r_d$, with
$r_d=r_s(z_d)$ and $z_d\simeq1059$.\\
For the same benchmark, with $\theta_s=\theta_s^{\Lambda}$ and $\lambda_1=0.15$, we obtain
at $z=0.5$
\be\label{eq3.26}
\frac{\Delta(D_M/r_d)}{D_M/r_d}\simeq-5.0\%, \qquad
\frac{\Delta(D_H/r_d)}{D_H/r_d}\simeq-2.26\%,
\ee
and for the isotropic distance
\be\label{eq3.27}
D_V(z)=\left[D_M^2(z)\frac{cz}{H(z)}\right]^{1/3},
\ee
the deviation is
\be\label{eq3.28}
\frac{\Delta(D_V/r_d)}{D_V/r_d}\simeq -4.22\%
\ee
Hence, the solution $H_0\simeq 74.7$ that reproduces $\theta_s$ exactly does not reproduce the BAO distances. 
If we demand simultaneously
\be\label{eq3.29}
\frac{r_s}{D_M(z_{*})}=\frac{r^{\Lambda}_s}{D^{\Lambda}_M(z_{*})}
\ee
and
\be\label{eq3.30}
\frac{D_M(0.5)}{r_d}=\frac{D_M^{\Lambda}(0.5)}{r^{\Lambda}_d},
\ee
leaving $h$ and $\lambda_1$ free, then the solution is found at 
$h\simeq 0.671$ and $\lambda_1 \simeq -0.502$, that is $H_0\simeq 67.1~{\rm km\,s^{-1}\,Mpc^{-1}}$, and the root is no longer phantom. Therefore, an exact fit of both the CMB acoustic angle and the
reference BAO distance practically removes the amplification of $H_0$.\\
A more realistic approach is to allow small deviations in the BAO distances. We maximize
$h$ over $(\lambda_1,\lambda_2,\omega_2)$ requiring
\be\label{eq3.31}
\left|\frac{\Delta(D_M/r_d)}{D_M/r_d}\right|<\epsilon, \qquad
\left|\frac{\Delta(D_H/r_d)}{D_H/r_d}\right|<\epsilon, \qquad \epsilon=1\%,\,2\%,
\ee
together with
\begin{equation}
\theta_s=\theta_s^{\Lambda},\qquad f_{\cal H}(z_{eq})\le5\%,\qquad f_{\cal H}(z_{BBN})<1\%.
\end{equation}
For a BAO tolerance $\epsilon=1\%$ the maximum is reached at
$\lambda_1\simeq0.2995$, $\lambda_2\simeq-3.094$ and $\omega_2\simeq3.51\times10^{-3}$,
which gives
\begin{equation}
H_0^{\max}\simeq70.5~{\rm km\,s^{-1}\,Mpc^{-1}}, \qquad f_{\cal H}(z_{eq})\simeq2.57\%.
\end{equation}
In this case the transverse condition is saturated,
$\Delta(D_M/r_d)/(D_M/r_d)\simeq-1.00\%$, while the radial one remains well inside the
tolerance, $\Delta(D_H/r_d)/(D_H/r_d)\simeq0.53\%$.\\

\noindent For $\epsilon=2\%$ the maximum moves to $\lambda_1\simeq0.2994$, $\lambda_2\simeq-3.138$
and $\omega_2\simeq2.75\times10^{-3}$, with
\begin{equation}
H_0^{\max}\simeq71.6~{\rm km\,s^{-1}\,Mpc^{-1}}, \qquad f_{\cal H}(z_{eq})\simeq2.87\%,
\end{equation}
and $\Delta(D_M/r_d)/(D_M/r_d)\simeq-2.00\%$,
$\Delta(D_H/r_d)/(D_H/r_d)\simeq-0.08\%$.
In both cases the radial BAO distance is practically unaffected, and it is the transverse
distance that limits the increase of $H_0$. BBN does not impose any relevant restriction,
since $f_{\cal H}(z_{BBN})\lesssim10^{-6}$.

\subsection*{Two early modes}
We now include the $\lambda_3$ mode. Given $\lambda_1$ and $\lambda_2$, the third root is
fixed by the characteristic equation,
\be\label{lambda3}
\lambda_3=-\frac{(\lambda_1+7)\lambda_2+7\lambda_1+37}{\lambda_1+\lambda_2+7}.
\ee
As discussed in Section~3, the two early modes displace the effective matter-radiation
equality in opposite directions, so that the inclusion of $\lambda_3$ allows to control this
displacement. We maximize $h$ over $(\lambda_1,\lambda_2,\lambda_3,\omega_2,\omega_3)$,
subject to the conditions (\ref{eq3.31}) and (\ref{lambda3}), together with 
\begin{equation}
\theta_s=\theta_s^{\Lambda},\qquad
\frac{|\Delta z_{eq}|}{z_{eq}}\le5\%,\qquad
f_{\cal H}(z_{eq})\le5\%,\qquad
f_{\cal H}(z_{BBN})<1\%,
\end{equation}
for $\lambda_1,\;\lambda_2,\;\lambda_3$ within the ranges
\begin{equation}
0.15\le\lambda_1\le0.30,\qquad -3.53\le\lambda_2<-3,\qquad -4<\lambda_3\le-3.53 .
\end{equation}
In what follows we use the exact value $z_{eq}\simeq3422.9$.
For a BAO tolerance $\epsilon=1\%$ the maximum is found at
\begin{equation}
\lambda_1=0.300,\qquad \lambda_2\simeq-3.0679,\qquad \lambda_3\simeq-3.9471,
\end{equation}
with physical amplitudes $\omega_2\simeq6.5\times10^{-3}$ and
$\omega_3\simeq1.26\times10^{-6}$. For this configuration
$f_{\cal H}(z_{eq})\simeq4.7\%$, $\Delta z_{eq}/z_{eq}=5.00\%$,
$f_{\cal H}(z_{BBN})\simeq1.00\%$ and $\Delta r_d/r_d\simeq-2.315\%$, while the BAO
deviations are $\Delta(D_M/r_d)/(D_M/r_d)=-1.00\%$ and
$\Delta(D_H/r_d)/(D_H/r_d)=0.55\%$. The resulting maximum is
\begin{equation}
H_0^{\max}\simeq71.29~{\rm km\,s^{-1}\,Mpc^{-1}}.
\end{equation}

\noindent For $\epsilon=2\%$ we obtain
\begin{equation}
\lambda_1=0.300,\qquad \lambda_2\simeq-3.0922,\qquad \lambda_3\simeq-3.9277,
\end{equation}
with $\omega_2\simeq5.68\times10^{-3}$ and $\omega_3\simeq1.59\times10^{-6}$. In this case
$f_{\cal H}(z_{eq})\simeq5\%$, $\Delta z_{eq}/z_{eq}=5.00\%$,
$f_{\cal H}(z_{BBN})\simeq0.844\%$, $\Delta r_d/r_d\simeq-2.448\%$,
$\Delta(D_M/r_d)/(D_M/r_d)=-2.00\%$ and $\Delta(D_H/r_d)/(D_H/r_d)=-0.057\%$, which gives
\begin{equation}
H_0^{\max}\simeq72.41~{\rm km\,s^{-1}\,Mpc^{-1}}.
\end{equation}

\noindent The results for one and two early modes are compared in Table~VI.

\begin{center}
\begin{tabular}{lcccc}\hline\hline
 ${\bf Configuration}$  \hspace{0.9cm} & ${\bf BAO}$  \hspace{0.9cm}  & $H_0^{max}$\\ 
\hline
One early mode & $1\%$ & $70.5$ \\
Two early modes & $1\%$ & ${\bf 71.3}$ \\ 
One early mode & $2\%$ & $ 71.6$ \\
Two early modes & $2\%$ & ${\bf 72.4}$ \\ 
 \hline
\end{tabular}
\end{center}
\begin{center}
{\bf Table VI}
\end{center}

\noindent For both tolerances the inclusion of the $\lambda_3$ mode increases the geometric maximum
by approximately $\Delta H_0\simeq0.8~{\rm km\,s^{-1}\,Mpc^{-1}}$. In these solutions the
near-matter mode $\lambda_2$ provides most of the reduction of the sound horizon, while the
near-radiation mode $\lambda_3$ helps to keep the shift of the effective equality under
control, which allows a larger amplitude of the $\lambda_2$ mode than in the single-mode
case. As in the previous section, the transverse BAO distance is the observable that limits
the amplification of $H_0$.

\section{Discussion}
We have introduced an extended local infrared cutoff for the holographic density constructed as a derivative expansion up to third order of local quantities associated with the Hubble parameter. By requiring the cosmological equation to remain linear in $E=H^2$, the resulting background equation admits exact analytical solutions. It was found that two of the model parameters can be constrained so that the matter and radiation-like residual terms of the original local construction are exactly eliminated. Thus, the standard cosmological evolution of matter and radiation is recovered exactly, while the new physical effect is encoded exclusively in the solution of the third-order homogeneous equation.
This provides an exact separation between the standard cosmological fluids and the holographic sector and leads to a broader class of dynamical dark-energy solutions. The model therefore provides a framework for studying both late-time accelerated expansion and possible departures from the standard expansion history at earlier epochs.\\
\noindent The roots of the characteristic equation give rise to several cosmologically interesting branches. In particular, two negative roots may lie between the matter and radiation scalings. 
Their contributions grow faster than matter but slower than radiation toward the past, allowing them to modify the expansion rate around recombination and matter-radiation equality while remaining strongly suppressed at BBN. At the background level, a single near-matter mode can reduce the sound horizon by several percent. We considered some configurations that reduce the sound horizon in approximately $4.6\%-4.7\%$, although these require an additional contribution close to ($10\%$) around equality and should be regarded as upper geometrical limits.\\
\noindent When a slightly positive characteristic root, corresponding to an effective phantom component is considered, then the late-time distance-redshift relation is modified, which introduces an additional degeneracy between the acoustic scale and $H_0$. Fixing the CMB acoustic angle alone can therefore produce very large values of $H_0$, but such solutions are strongly constrained by BAO distances. When representative BAO deviations are restricted to the ($1\%-2\%$) level, the allowed background solutions typically give
$$
H_0\simeq70.5-71.6\ {\rm km\;s^{-1}\;Mpc^{-1}}
$$
for a single early mode.\\
\noindent When the second early mode is included, it can reduce the displacement of the effective matter-radiation equality. This is due to the fact that the two negative modes lie on opposite sides of the effective ($w=1/6$) scaling and their contributions to the displacement of effective matter-radiation equality can partially compensate each other, while both continue to increase the pre-recombination expansion rate and reduce the sound horizon.
Optimizing  the three-mode configurations leads to the following approximations
$$
H_0\simeq71.3\ {\rm km\;s^{-1}\;Mpc^{-1}}
$$
for a $1\%$ BAO tolerance, and values around
$$
H_0\simeq72.4\ {\rm km\;s^{-1}\;Mpc^{-1}}
$$
for a ($2\%$) tolerance. In these solutions the near-matter mode provides most of the acoustic-horizon reduction, while the near-radiation mode helps control the shift of effective equality.\\
\noindent These results show that the third-order homogeneous modes generated by the extended holographic cutoff can produce a viable late-time cosmological expansion and a non-negligible reduction of the pre-recombination sound horizon, without affecting the expansion rate during BBN.         
The present analysis is restricted to the homogeneous background, and the quoted values of $H_0$ should therefore be considered as estimates of the geometrical potential of the model to address the \(H_0\) tension. A complete study requires cosmological perturbations and statistical comparison with CMB, BAO, supernovae and large-scale-structure observations.
The main purpose of the present work is instead to establish the extended holographic cutoff and analyze its background dynamics. This model extends the scope of holographic cosmology and enriches the variety of late and early-time cosmological scenarios.

\section*{Acknowledgments}
This work was supported by Universidad del Valle under project CI 71436.

\end{document}